\documentclass[fleqn,12pt,twoside]{article}
\usepackage[headings]{espcrc1}
\usepackage{t1enc,wrapfig}

\usepackage{graphicx}

\usepackage[figuresright]{rotating}

\title{Near-barrier Fusion and Transfer/Breakup induced by Weakly Bound
and Exotic Halo Nuclei}

\author{C. Beck\address {Institut Pluridisciplinaire Hubert Curien, UMR7178,
IN2P3-CNRS et Universit\'{e} Louis Pasteur (Strasbourg I), 23 rue du Loess -
BP28, F-67037 Strasbourg Cedex 2, France} }

\runtitle{Near-barrier Fusion and Breakup/Transfer induced by Weakly
Bound Nuclei} \runauthor{C. Beck}

\begin{document}

\maketitle

\begin{abstract}

{The influence on the fusion process of coupling to collective
degrees of freedom has been explored. The significant enhancement of
the fusion cross section at sub-barrier energies was compared to
predictions of one-dimensional barrier penetration models. This was
understood in terms of the dynamical processes arising from strong
couplings to collective inelastic excitations of the target and
projectile. However, in the case of reactions where at least one of
the colliding nuclei has a sufficiently low binding energy, for
breakup to become an important process, conflicting model
predictions and experimental results have been reported in the
literature. Excitation functions for sub- and near-barrier total
(complete + incomplete) fusion cross sections have been measured for
the $^{6,7}$Li+$^{59}$Co reactions. Elastic scattering as well as
breakup/transfer yields have also been measured at several incident
energies. Results of Continuum-Discretized Coupled-Channel ({\sc
Cdcc}) calculations describe reasonably well the experimental data
for both reactions at and above the barrier. A systematic study of
$^{4,6}$He induced fusion reactions with a three-body {\sc Cdcc}
method is presented. The relative importance of breakup and
bound-state structure effects on total fusion (excitation functions)
is particularly investigated. The four-body {\sc Cdcc} model is
being currently developed. }

\end{abstract}

\section{INTRODUCTION}

In reactions induced by weakly bound nuclei, the influence on the fusion
process of couplings both to collective degrees of freedom and to
breakup/transfer channels is a key point for the understanding of N-body
systems in quantum dynamics. Due to the very weak binding energies of halo
nuclei, such as $^{6}$He or $^{11}$Be \cite{Hagino00,Raabe04,Liang06,Canto06},
a diffuse cloud of neutrons would lead to enhance fusion probabilities below
the Coulomb barrier, where the neutron tail which extends well beyond the
compact nuclear core provides a conduit by which the matter distributions of
the target and projectile may overlap at longer range than for the core. In
vicinity of the Coulomb barrier and below, the enhanced fusion with $^{11}$Be
was predicted~\cite{Hagino00}. On the other, the main enhancement effect
for fusion with $^{6}$He compared to its $^{4}$He core may be due to the
neutron rearrangement giving a gain in energy~\cite{Zagrebaev03} as confirmed
experimentally with $^{206}$Pb and $^{197}$Au targets~\cite{Penion06}. However,
recent experimental studies involving $^{6}$He radioactive ion beams (RIB)
\cite{Raabe04,Kolata98,Bychowski04,Dipietro04,Navin04,DeYoung05} indicate that
the halo of the $^{6}$He nucleus does not enhance the fusion probability,
illustrating the preponderant role of one- and two-neutron transfers (TR) in
$^{6}$He induced fusion reactions. Hence, the question of a real new effect
with RIB's and with stable beams such as weakly bound $^{6}$Li, $^{7}$Li and
$^{9}$Be nuclei, namely the occurence of non-conventional transfer/stripping
processes with large cross sections most likely originating from the small
binding energy of the projectile, remains open~\cite{Liang06,Canto06}.

     \begin{wrapfigure}[32]{r}[0cm]{8cm}
\includegraphics[width=8cm,height=12cm]{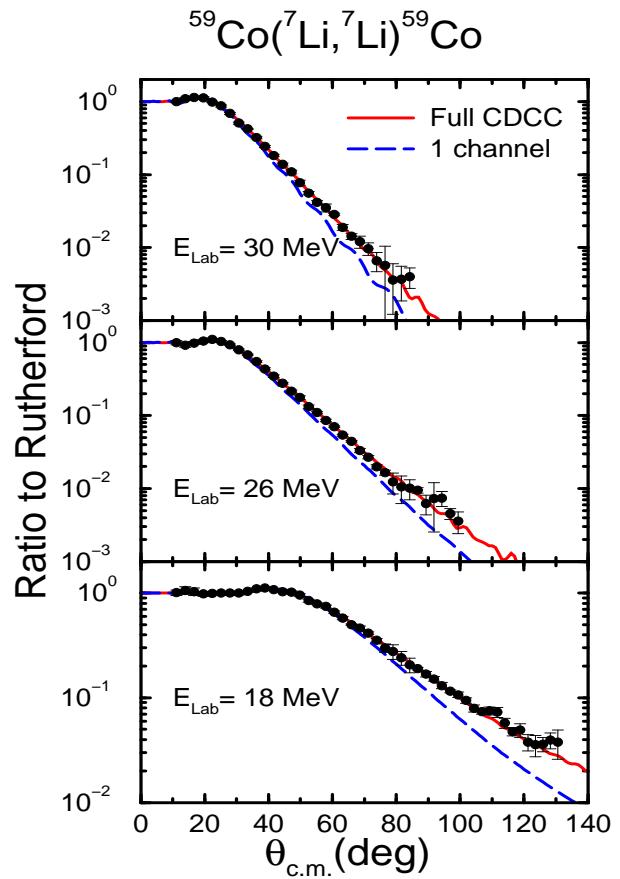}       \parbox{80mm}
    {\caption{\label{}{\small Experimental elastic scattering (data
                      points) for
$^{7}$Li+$^{59}$Co~\cite{Beck04}. The theoretical
curves correspond to {\sc Cdcc}
calculations~\cite{Beck06} with (solid lines) or without
       (dashed lines) couplings with the continuum
 as discussed in the text.}}}        \end{wrapfigure}

Since the coupling between channels is known to enhance the fusion cross
section at sub-barrier energies~\cite{Hagino00,Liang06,Canto06},
coupled-channel (CC) effects have to be taken into account in the theoretical
description of the fusion process. A large number of experimental results have
been interpreted adequately well within the framework of CC calculations.
However, in the case of loosely bound (and/or halo) systems the situation is
more complicated since the breakup channel couples strongly to an infinite
number of unbound states into the continuum~\cite{Hagino00,Canto06,Diaz02}. A
possible treatment of the problem is to reduce it to a finite number of
channels. The traditional approach to discuss the sub-barrier fusion reaction
induced by weakly bound nuclei is to solve the CC equations by discretizing in
energy the particle continuum states in the projectile
nucleus~\cite{Hagino00,Diaz02}. This is the so-called method of
Continuum-Discretized Coupled-Channels ({\sc Cdcc}) that has been initially
proposed by Rawitscher \cite{Rawitscher74}, and later developed for light
heavy-ion reactions~\cite{Austern87}. Experimental results~\cite{Tripathi02}
have well confirmed the calculations that fusion cross sections are
significantly enhanced due to the couplings to the continuum states at energies
below the barrier, while they are hindered above~\cite{Hagino00}. With the
recent advent of new RIB facilities
\cite{Liang06,Penion06,Kolata98,Dipietro04,Navin04}, it should be now possible
to constrain more the parameters of the {\sc Cdcc} formalism~\cite{Diaz02}.
Several studies have been initiated in this direction
\cite{Rusek03,Keeley03,Rusek05,Beck04,Beck06} to investigate reactions induced
by the neutron halo of the borromean $^{6}$He nucleus, which is known to have a
strong dipole excitation mode.

In the recent past a {\sc Cdcc} calculation~\cite{Rusek05} failed to reproduce
the large yields of $\alpha$-particles reported in the $^{6}$He+$^{209}$Bi
reaction~\cite{Kolata98}. In this work we propose full {\sc Cdcc} calculations
describing simultaneously elastic scattering (see Fig.~1), total fusion, and
breakup of weakly bound light nuclei both stable ($^{6}$Li and $^{7}$Li) and
radioactive ($^{6}$He) with a medium-mass target ($^{59}$Co). Preliminary brief
reports have been given elsewhere~\cite{Beck04,Beck06}.

\section{CONTINUUM-DISCRETIZED COUPLED-CHANNEL CALCULATIONS}

{\sc Cdcc}, a fully quantum-mechanical method developed originally to study the
effect of deuteron breakup on the process of elastic
scattering~\cite{Rawitscher74}, has been widely applied by the Kyushu
group~\cite{Austern87} to study heavy-ion collisions induced by light weakly
bound nuclei. {\sc Cdcc} calculations have been successful in the past in
describing the scattering of deuterons and $^{6,7}$Li~\cite{Austern87} on
different targets. The {\sc Cdcc} method has been then applied to reactions
with halo nuclei and, as a consequence, Diaz-Torres and Thompson \cite{Diaz02}
have been able to perform a full calculation of the theoretical fusion cross
section of halo nuclei using a novel method still based on the {\sc Cdcc}
formalism. A more recent study of the $^{6}$He+$^{209}$Bi reaction by means of
a three-body {\sc Cdcc} model~\cite{Rusek05} has shown much larger absorption
cross sections than experimental fusion cross sections~\cite{Kolata98}. In the
present work similar {\sc Cdcc} calculations are applied for the interaction of
$^{6,7}$Li and $^{6}$He (with the simultaneous description of elastic
scattering, fusion and breakup) with a medium-mass target such as $^{59}$Co,
and comparisons with the corresponding experimental data are presented.

Details of the calculations concerning the breakup space (number of partial
waves, resonances energies and widths, maximum continuum energy cutoff,
potentials, ...) have been given elsewhere~\cite{Diaz02,Diaz03}, in particular
in Tables I, II and III of Ref.~\cite{Diaz03}. The {\sc Cdcc} scheme is
available in the general coupled channels (CC) computer code {\sc
FRESCO}~\cite{Thompson88}. All calculations were carried out using the version
FRXX.09g of {\sc FRESCO} \cite{Thompson88}. Our choice was mainly influenced by
the fact that we have already carried out extended {\sc Cdcc} calculations for
both the $^{6}$Li+$^{59}$Co and $^{7}$Li+$^{59}$Co total fusion
reactions~\cite{Diaz03} which data were previously
published~\cite{Beck03,Szanto03,Souza04}. Before investigating that the
proposed {\sc Cdcc} formalism can be also applied to halo nuclei such as
$^{6}$He, we present the full description of the $^{6}$Li $\rightarrow$
$\alpha$+$d$ and $^{7}$Li $\rightarrow$ $\alpha$+$t$ clusters as two-body
objects, respectively, including elastic scattering angular distributions,
total fusion cross sections, and breakup cross sections.

We would like to stress that in the chosen fusion calculations the imaginary
components of the off-diagonal couplings in the transition potentials have been
neglected, while the diagonal couplings include imaginary parts \cite{Diaz03}.
Otherwise full continuum couplings have been taken into account so as to
reproduce elastic scattering data when available. We have used short-range
imaginary fusion potentials for each fragment separately ($\alpha$ and
$d$+target potentials and $\alpha$ and $t$+target potentials for $^{6}$Li and
$^{7}$Li, respectively). This is equivalent to the use of incoming wave
boundary conditions as performed in {\sc Ccfull} calculations~\cite{Beck03},
for instance.

\subsection{{\sc Cdcc} calculation of $^{7}$Li+$^{59}$Co and $^{6}$Li+$^{59}$Co
elastic scattering}

Results of the comparison of the {\sc Cdcc} calculations for the elastic
scattering with the data~\cite{Beck04,Souza04} are shown in Fig.~1 and Fig.~2
for $^{7}$Li+$^{59}$Co and $^{6}$Li+$^{59}$Co, respectively. The two different
curves are the results of calculations performed with (solid lines) and without
(dashed lines) $^{6,7}$Li $\rightarrow$ $\alpha$ + $d,t$ breakup couplings to
the continuum (i.e. continuum couplings). The agreement is very good when full
continuum couplings are taken into account. This was also found for the elastic
scattering of both $^{7}$Li+$^{65}$Cu and $^{6}$Li+$^{65}$Cu
reactions~\cite{Shrivastava06}. The effect of breakup on elastic scattering,
stronger for $^{6}$Li as expected, is illustrated by the difference between the
one-channel calculations (equivalent to the optical-model OM
calculations~\cite{Souza04}) and the full {\sc Cdcc} results. The {\sc Cdcc}
calculations using similar potentials that fit the measured elastic scattering
angular distributions~\cite{Souza04} of Fig.~2 are able to reproduce the
breakup cross sections measured for $^{6}$Li+$^{59}$Co~\cite{Beck06,Souza05}.
For $^{6}$Li the total calculated breakup cross sections, obtained by
integrating contributions from the states in the continuum up to 8 MeV, are
small as compared with total fusion {\sc Cdcc}~\cite{Diaz03} or
experimental~\cite{Beck03} (data points in the right panel of Fig.~3) cross
sections.

      \begin{wrapfigure}[32]{r}[0cm]{8cm}
\includegraphics[width=8cm,height=12cm]{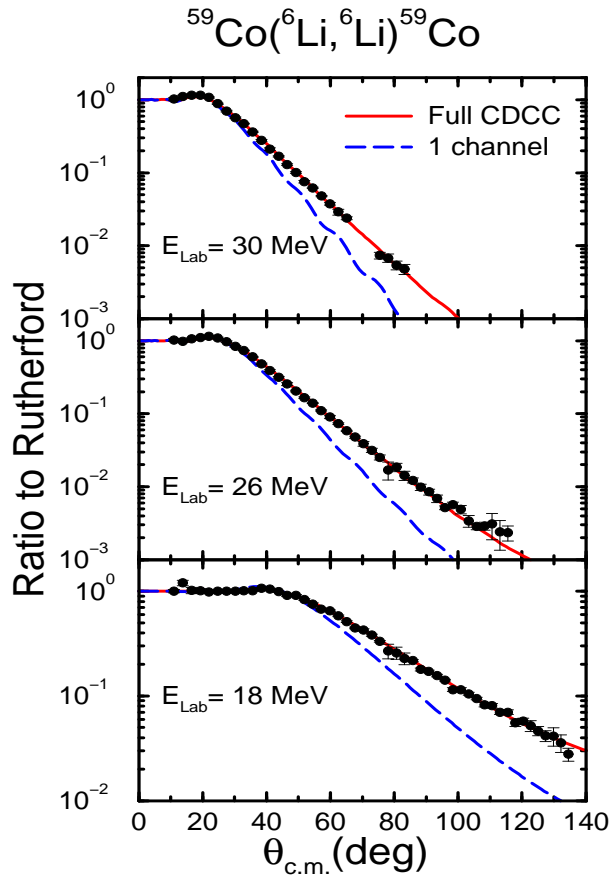}         \parbox{80mm}
      {\caption{\label{}{\small Experimental elastic scattering (data points)
                            for $^{6}$Li+$^{59}$Co~\cite{Beck04}.
                The theoretical curves correspond to {\sc Cdcc}
              calculations~\cite{Beck06} with (solid lines) or without
                     (dashed lines) couplings with the continuum
               as discussed in the text (see Sec.2.1).}}}
\end{wrapfigure}

The dynamic polarization potentials ({\sc Dpp}) generated by the coupling to
breakup channels are determined by the {\sc Cdcc} method. The real and
imaginary parts of the {\sc Dpp} are obtained at the three bombarding energies
for $^{6}$Li+$^{59}$Co and $^{7}$Li+$^{59}$Co; their energy dependences appear
to show the threshold anomaly only for the latter reaction. This preliminary
conclusion will need to be confirmed by subsequent measurements at several
other near-barrier energies.

\subsection{{\sc Cdcc} calculations of $^{6}$He+$^{59}$Co fusion}

In the following we present similar calculations applied for the two-neutron
halo nucleus $^{6}$He. The present case is much more complicated since $^{6}$He
breaks into three fragments ($\alpha$+n+n) instead of two ($\alpha$+d), and the
{\sc Cdcc} method has not yet been developed for two-nucleon halo nuclei. Hence
a di-neutron model is adopted for the $^{6}$He+$^{59}$Co reaction: i.e. we
assume a two-body cluster structure of $^{6}$He = $^{4}$He+$^{2}$n with an
$\alpha$-particle core coupled to a single particle, a di-neutron ($^{2}$n).
Couplings to resonant (2$^{+}$, E$_{ex}$ = 0.826 MeV) and non-resonant
continuum states (up to f-waves) are included. It is important to notice that
the fact that the di-neutron is not an object with both fixed size and fixed
energy (Heisenberg principle) might be a critical point in the present model.

The maximum energy of the continuum states is still equal to 8 MeV. Similarly
to our previous work~\cite{Diaz03}, the potentials between the fragments and
the $^{59}$Co target are those obtained with the global Broglia-Winther
Woods-Saxon parametrization~\cite{Broglia91}. (the numerical values are:
V$_{o}$ = -16.89 MeV, r$_{o}$ = 1.09 fm and a = 0.63 fm). For the $\alpha$-2n
binding potential (0$^{+}$, g.s.) we have used the following Woods-Saxon
potential: V$_{o}$ =  -40.796 MeV, r$_{o}$ = 1.896 fm and a = 0.3 fm. The g.s.
binding potential of the $\alpha$ particle and the di-neutron provides a 2s
bound state of about -0.975 MeV. The binding potential of the 2$^{+}$ resonant
state has also a Woods-Saxon form with the following parameters: V$_{o}$ =
-35.137 MeV, r$_{o}$ = 1.896 fm, a = 0.3 fm. With this potential the energy of
the 2$^{+}$ resonant state is 0.826 MeV and its width is 0.075 MeV. To obtain
converged (within a 5$\%$ level) total fusion cross section we have included:
(i) partial waves for $\alpha$-2n relative motion up to f-waves ($l$ = 3), (ii)
the $^{6}$He fragment-target potential multipoles up to the octupole term, and
(iii) the maximum of the continuum energy is 8 MeV. All resonant and
non-resonant continuum couplings including continuum-continuum couplings were
included in the calculation. Results of the {\sc Cdcc} calculations for the
fusion of $^{6}$He+$^{59}$Co are compared to the $^{4}$He+$^{59}$Co and
$^{6}$Li+$^{59}$Co systems in Fig.~3 and discussed below.

       \begin{figure}        \includegraphics[height=10cm]{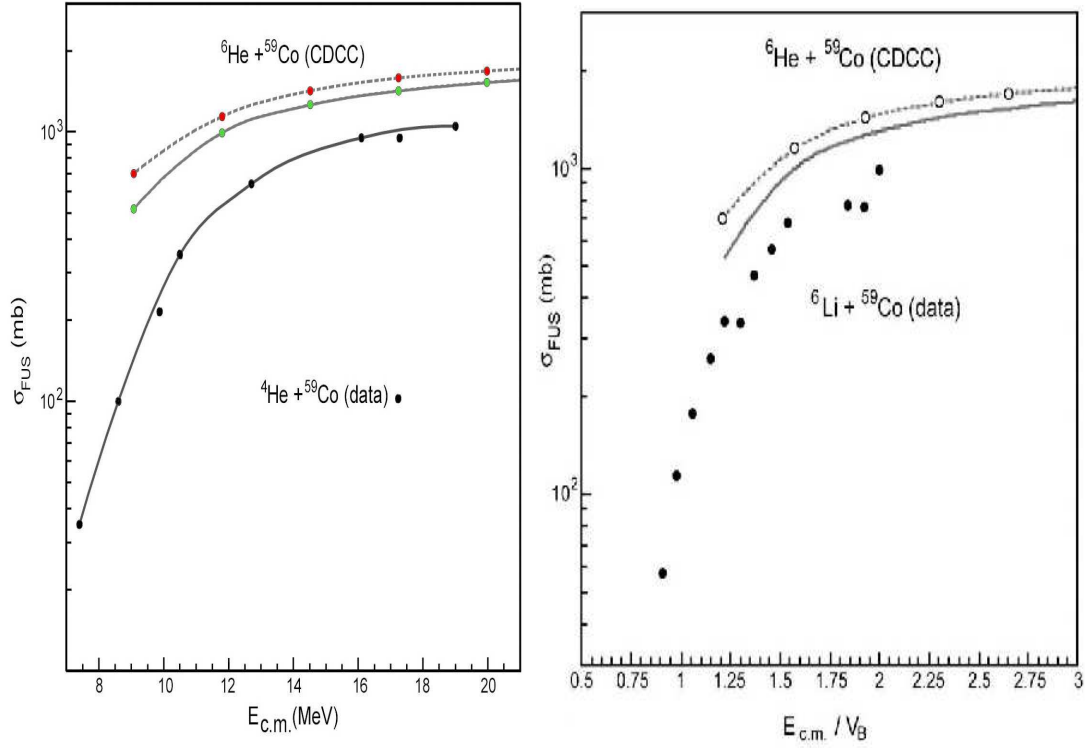}
                     {\caption{\label{}{\small Total fusion excitation
                     functions for $^{4}$He+$^{59}$Co (data points
                 \cite{McMahan80} and solid black line for
         {\sc CC} predictions on left panel) for
$^{6}$Li+$^{59}$Co (data points~\cite{Beck03}                               on
right panel), and for                               $^{6}$He+$^{59}$Co. The
curves correspond to                               {\sc Cdcc}
calculations~\cite{Diaz03} for $^{6}$He+$^{59}$Co
with (dashed line) or without (thin                               line)
couplings to the continuum as                               discussed in the
text (see Sec.2.2).}}}         \end{figure}

\section{DISCUSSION}

The proposed three-body {\sc Cdcc} model is adequate for $^{7}$Li as well as
for $^{11}$Be, as long as core excitation is ignored; and, probably better than
that for $^{6}$Li. However the elastic scattering data \cite{Souza04} as
plotted in Figs.~1~and~2 are very well reproduced for both $^{7}$Li and
$^{6}$Li. It is interesting to note that the OM analysis \cite{Souza04} of
their respective angular distributions was found to be ambiguous for the two
lowest energies when using the parameter-free S\~ao Paulo OM
potential~\cite{Chamon02}. The extracted total reaction cross sections
\cite{Souza04} confirm the observed small enhancement ($\approx$ factor 2) of
total fusion cross section for the more weakly bound $^{6}$Li nucleus at
sub-barrier energies~\cite{Beck03}, with similar yields for both reactions at
and above the Coulomb barrier in concordance with {\sc Cdcc}
calculations~\cite{Diaz03}. This moderate enhancement effect observed below the
barrier can be explained by the fact that both
experimental~\cite{Beck06,Souza05} and theoretical breakup cross
sections~\cite{Beck06} are rather small (10-50 mb).

Without available data for $^{6}$He+$^{59}$Co we have cautiously decided not to
present {\sc Cdcc} calculations for the elastic scattering as the Coulomb
dipole excitation is known to be too strong in the di-neutron
approximation~\cite{Rusek05}, and then difficult to be predicted. The dipole
Coulomb excitation of $^{6}$He projectiles in the field of a highly charged
target has already been discussed \cite{Rusek03,Keeley03,Rusek05}. The
di-neutron {\sc Cdcc} model works much better when the dipole coupling strengh
is reduced by 50 $\%$. This reduction is important for heavy targets, but
probably not as much for a medium-mass target such as $^{59}$Co. Nevertheless,
such reduction is also reducing total absorption cross section in the {\sc
Cdcc} calculations. If we consider this cross section as the total fusion cross
section, we may overestimate the fusion of $^{6}$He+$^{59}$Co slightly.
Obviously, the search for similar effects is of high interest for this
medium-mass region~\cite{Dipietro04}, and elastic measurements with higher
precision will have to be undertaken.

There is some confusion about the definition of fusion~\cite{Diaz03}. Theorists
usually define complete fusion (CF) as the capture of all projectile fragments,
and incomplete fusion (ICF) the capture of only some fragments~\cite{Canto06}.
Among other CC calculations, {\sc Cdcc} has the disadvantage of being unable to
distinguish between CF and ICF~\cite{Diaz03}. In experiments a similar
complication arises from the lack of a clear separation of CF and ICF cross
sections~\cite{Beck03}: therefore CF is defined experimentally as the capture
of all the charge of the projectile by the target
\cite{Beck03,Dasgupta04,Liu05}. It is believed that there is a
significant contribution of breakup followed by CF to the total fusion cross
sections~\cite{Canto06}. The combined effect of breakup and TR in the CC
({\sc Cdcc}) approach has not been fully studied so far in the context of
sub-barrier fusion.

{\sc Cdcc} calculations using the set of parameters given for the
$^{6}$He+$^{59}$Co reaction in the previous Section are displayed in Fig.~3.
The present calculations do not include neither target excitations nor
TR channnels. However with crude estimations as those performed for the
$^{6}$Li+$^{59}$Co reaction~\cite{Diaz03} the effect is found to be very small.
In Fig.~3 we compare the total fusion excitation functions of the two
$^{6}$He+$^{59}$Co ({\sc Cdcc} calculations) and $^{6}$Li+$^{59}$Co
(experimental data of Ref.~\cite{Beck03}) reactions. For the $^{6}$He reaction,
the incident energy is also normalized with the Coulomb barrier V$_{B}$ of the
bare potential. The first calculation (solid line) only include the
reorientation couplings in fusion without breakup. All contiunum and
reorientation couplings are included in fusion with breakup (dashed curve). In
agreement with heavier targets data~\cite{Penion06}, we can observe that both
calculated curves (with and without breakup) give much larger total fusion
cross section for $^{6}$He as compared to $^{6}$Li. We can also observe that
the inclusion of the couplings to the breakup channels notably increases the
total fusion cross section for the whole energy range. Similar conclusions are
reached when $^{6}$He+$^{59}$Co ({\sc Cdcc} calculations) is compared to
$^{4}$He+$^{59}$Co (here the {\sc Cdcc} calculations are fitting the data of
Ref.~\cite{McMahan80} remarkably well) in Fig.~3.

\section{SUMMARY AND OUTLOOK}

Halo and cluster nuclei, with well-defined breakup and fusion modes, are good
test-benches for theories of fusion and breakup. A more complete theoretical
model of few-body dynamics that is able to distinguish CF from ICF will need to
follow correlations after breakup, so we need either three-body (and most
preferably four-body~\cite{Moro06}) state-of-art {\sc Cdcc}
\cite{Diaz02,Rusek03,Keeley03,Diaz03} calculations of the type we have
presented here as a starting point or, for instance, the use of a
time-dependent wave-packet model~\cite{Ito06}.

The {\sc Cdcc} formalism, with continuum-continuum couplings taken into
account, is most probably one of the most accurate method nowadays. Therefore,
a systematic study of $^{4,6}$He induced fusion reactions with the {\sc Cdcc}
method is being undertaken. However up to now only very scarce studies with
$^{6}$He projectiles are available
\cite{Raabe04,Penion06,Kolata98,Bychowski04,Dipietro04,Navin04}: data from {\sc
Spiral} and Louvain-la-Neuve have recently been published for
$^{6}$He+$^{63,65}$Cu~\cite{Navin04} and for $^{6}$He+$^{64}$Zn
\cite{Dipietro04}. The present extensive {\sc Cdcc} calculations show that for
$^{6}$He+$^{59}$Co considerable enhancement of the sub-barrier fusion cross
sections is predicted as compared to measured fusion yields for both the
$^{6}$Li+$^{59}$Co~\cite{Beck03} and $^{4}$He+$^{59}$Co~\cite{McMahan80}
systems. This conclusion is consistent with Dubna data for heavier
targets~\cite{Penion06}. We should note that contradictory theoretical results
have been obtained for $^{10,11}$Be+$^{209}$Bi collisions with a different
approach based on the time-dependent wave-packet formalism~\cite{Ito06}.

The {\sc Cdcc} method~\cite{Diaz03}, which is shown here to be quite succesfull
for fusion with stable nuclei, will be used to provide the complete theoretical
description of all competing processes (total fusion, elastic scattering, and
breakup) in a consistent way. One really needs to investigate such processes in
the dynamics of the interaction at the Coulomb barrier with loosely bound halo
nuclei. The undertanding of the reaction dynamics involving couplings to the
breakup channels requires the explicit measurement of precise elastic
scattering data as well as yields leading to the TR (although TR cannot be
described within {\sc Cdcc}) and breakup (and/or ICF) channels. The complexity
of such reactions, whereby many processes compete on an equal footing,
necessitates kinematically and spectroscopically complete measurements, i.e.,
ones in which all processes from elastic scattering to CF are measured
simultaneously, providing a technical challenge in the design of broad range
detection systems. A new experimental programme with {\sc Spiral} beams and
medium-mass targets is underway at {\sc Ganil} within the forthcoming
years~\cite{Navin04,Beck04,Beck06}.

While $^{6}$He is best described as a three-body $\alpha$-$n$-$n$ object, at
present the {\sc Cdcc} method has not yet been completely implemented for
four-body breakup~\cite{Moro06}, its two-body $\alpha$-$^{2}$n model appears to
be rather satisfactory. The application of four-body {\sc Cdcc} models under
current development~\cite{Moro06,Matsumoto06} will then be highly desirable. In
the longer term, four-body {\sc Cdcc} models might also be required for
$^{6}$Li, due to the possible sequential breakup of the deuteron. The questions
in the theory of a halo system such as the borromean $^{6}$He nucleus, its
breakup (and in the breakup of many-body projectiles generally), and its CF and
ICF processes will need the knowledge not just of those integrated cross
sections, but the phase space distributions of the surviving fragment(s).
Therefore, future very exclusive experiments performed at sub-barrier energies
will have to determine very precisely the angular correlations of the
individual neutrons. Preliminary attempts of
measurements~\cite{Bychowski04,DeYoung05} of $\alpha$-particles in coincidence
with neutrons are promising to disentangle the effect of halo structures
on the reaction mechanisms.\\

\noindent
{\small {\bf Aknowledgments:} I would like to acknowledge N. Keeley and A.
Diaz-Torres for their collaboration in performing the {\sc Cdcc} calculations.
F.A. Souza is thanked for providing us with experimental data on breakup (cited
in Refs.~\cite{Beck06,Souza05}) prior to their publication. } \\

\end{document}